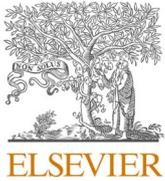
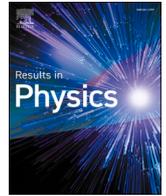

# Fermionic decays of NMSSM Higgs bosons under LHC 13 TeV constraints

Marwa Telba [*,1], Maien Binjonaid [1]

*Department of Physics and Astronomy, King Saud University, Riyadh, Saudi Arabia*

A B S T R A C T

This paper investigates the impact of the recent LHC constraints on the Higgs sector in the semi-constrained version of the Next-to-Minimal Supersymmetric Standard Model. Our analysis focuses on the parameter space for which the value of the Higgs doublet-singlet coupling, $\lambda$, is large as possible, while the ratio between the vacuum expectation values of the two Higgs doublets, $\tan\beta$, is small as possible. Under the current constraints, we present the possible fermionic decay channels and reduced cross-section into fermions final states for the lightest neutral Higgs bosons in the NMSSM, ($h_1, h_2$, and $a_1$). We found that the branching ratios of the non SM-like Higgs ($a_1$ and $h_2$) into a pair of bottom quarks are near 90% level when the Higgs mass below 400 GeV. Moreover, the branching ratio of $h_2/a_1 \rightarrow t\bar{t}$ can reach unity for all mass ranges when these bosons are mostly singlet.

## Introduction

In 2012, ATLAS and CMS at the Large Hadron Collider announced the detection of a Higgs-like particle with a mass of around 125 GeV and its measurements consistent with the Standard Model's prediction [1,2]. However, several new physics models with extended Higgs sector can also accommodate 125 GeV Higgs boson, such as the Minimal Supersymmetric Standard Model (MSSM) [3,4]. The Higgs sector in the MSSM involves two Higgs doublets, which the mixing of their components leads to five Higgs states, one of them is the SM-like Higgs. To obtain a 125 GeV Higgs, the MSSM needs large quantum corrections from the third generation of squark loops [5,6]. Also, in the MSSM Lagrangian, there is a supersymmetric term, the $\mu$-term. This term should be of a scale of the electroweak scale to ensure that the two Higgs doublets get a non-zero vacuum after EWSB; this leads to the well-known "$\mu$ problem" [7]. However, in the Next-to-minimal supersymmetric standard model (NMSSM) [8,9], these issues can be fixed. This model was proposed as an extension of MSSM with an additional Higgs singlet field, $\widehat{S}$, that generates a $\mu$ parameter dynamically of the SUSY breaking scale, solving the "$\mu$-problem" [10,11]. Moreover, this new singlet scalar gives the NMSSM particle spectrum additional degrees of freedom. In the Higgs sector of the NMSSM, there are seven massive states: three CP-even $h_i$ ($i = 1,2,3$), two CP-odd $a_j$ ($j = 1,2$), two charged $h^\pm$. Furthermore, the introduction of an extra Higgs singlet in the NMSSM has many implications that cause interesting phenomena. Thus, the NMSSM was widely proposed to interpret the results of the LHC [12–34]. The singlet components' presence causes all Higgs bosons measurements to deviate from the Higgs boson's expected values in the SM, which may change the particle (and sparticle) decay width and signals at the LHC [35,36]. Based on this observation, the constraints of the recent LHC Higgs data on the lightest neutral Higgs states' properties are discussed in this paper. This method was used in a previous paper to investigate the effect of the latest LHC constraints on bosonic decays of possible lightest neutral Higgs bosons in the semi-constrained NMSSM (scNMSSM) with Grand unification boundary conditions [37] and will be shortly described in Section "Scan strategy". In this paper, we will apply this method to predict the effect of the recent constraints of the LHC on fermionic decays of neutral Higgs particles in scNMSSM.

The structure of this paper is as follows. Section "Higgs Sector in the NMSSM" briefly introduces the Higgs sector in the NMSSM model. Section "Scan Strategy" is devoted to present the analyzed parameter space of the NMSSM at the GUT scale and the imposed theoretical and experimental constraints. In Section "Results", our results are derived and analyzed. Finally, Section "Conclusion" contains brief conclusions of our findings.

## Higgs sector in the NMSSM

The NMSSM represents the simplest extension of the MSSM by additional gauge chiral superfield which is singlet under $SU(3)_c \times SU(2)_L \times U(1)_Y$. To solve the $\mu$-problem of the MSSM, $Z_3$-symmetry is imposed. Thus, the scale-invariant NMSSM superpotential reads [8],

$$W_{NMSSM} = W_{Yukawa} + \lambda \widehat{S} \widehat{H}_u \cdot \widehat{H}_d + \frac{1}{3} \kappa \widehat{S}^3. \quad (1)$$

The first line in Eq. (1) represents the Yukawa couplings of the Higgs doublet fields ($\widehat{H}_d$ and $\widehat{H}_u$) to the lepton and quark superfields. On the other hand, the Higgs mass term in the MSSM is replaced by a linear coupling of $\widehat{S}$ to $\widehat{H}_d$ and $\widehat{H}_u$ plus the self-coupling term ($\frac{1}{3} \kappa \widehat{S}^3$). Once the singlet superfield gets a vacuum expectation value $\langle S \rangle = s$, the second term in NMSSM superfield generates an efficient $\mu$-term; to solve the $\mu$-problem [8].

$$\mu_{eff} = \lambda s. \quad (2)$$

The most general NMSSM soft Lagrangian is consists of mass terms for all scalars, the Higgs and sfermion fields ($m_{H_u}^2, m_{H_d}^2, m_S^2, m_Q^2, m_U^2, m_D^2, m_L^2$ and $m_E^2$), the gauginos mass terms ($m_1, m_2$ and $m_3$), and finally, the trilinear soft SUSY breaking interaction between the sfermions and Higgs fields.

$$-\mathscr{L}_{soft} = m_{H_u}^2 |H_u|^2 + m_{H_d}^2 |H_d|^2 + m_S^2 |S|^2 + m_Q^2 |Q|^2 + m_U^2 |U_R^2| + m_D^2 |D_R^2| + m_L^2 |L^2| + m_E^2 |E_R^2| + \frac{1}{2}\left[m_1 \lambda_1 \lambda_1 + m_2 \sum_{i=1}^{3} \lambda_2^i \lambda_{i2} + m_3 \sum_{a=1}^{8} \lambda_3^a \lambda_{a3}\right]$$
$$+ h_u A_u Q \cdot H_u U_R^2 - h_d A_d Q \cdot H_d D_R^2 - h_e A_e L \cdot H_d E_R^2 + \lambda A_\lambda H_u \cdot H_d S + \frac{1}{3} \kappa A_\kappa S^3 + h.c. \quad (3)$$

To present the tree-level mass matrices of the Higgs fields physically, the expansion of the full scalar potential around the vacuum expectation values (VEVs), $v_d, v_u$, and $s$, is required. Thus, the neutral Higgs doublets and singlet components are labeled by,

$$H_u^0 = v_u + \frac{1}{\sqrt{2}}(R_u^0 + iI_u^0), H_d^0 = v_d + \frac{1}{\sqrt{2}}(R_d^0 + iI_d^0),$$
$$S = s + \frac{1}{\sqrt{2}}(R_s + iI_s). \quad (4)$$

The CP-even Higgs bosons ($h_1, h_2, h_3$) are acquired from the mixing between a real part of $S$ with real parts of $H_d^0$ and $H_u^0$, while the CP-odd Higgs states ($a_1, a_2, a_3$) are obtained from the mixing between the imaginary parts of Higgs fields.

One of the advantages of this model is the prediction of the Higgs-like particle, at tree level, with mass around 125 GeV, where the following expression gives the upper bound on the lightest CP-even Higgs mass,

$$m_h^2 \leqslant M_Z^2 \left( \cos^2(2\beta) + \frac{\lambda^2}{g^2} \sin^2(2\beta) \right), \quad (5)$$

The maximum tree-level enchantment is achieved when the ratio between the VEVs of the two Higgs doublets, $\tan\beta \equiv v_u/v_d$, is small, and the coupling constant between the singlet and two Higgs doublet fields, $\lambda$, is large.

In scNMSSM, we assume that the soft SUSY breaking terms (the gauginos mass, sfermions mass, and the trilinear couplings) are equal at the GUT scale.

$$m_1 = m_2 = m_3 \equiv m_{1/2},$$
$$m_Q^2 = m_U^2 = m_D^2 = m_L^2 = m_E^2 \equiv m_0^2, \quad (6)$$
$$A_t = A_b = A_\tau \equiv A_0.$$

In addition, the Higgs soft mass terms $m_{H_d}^2, m_{H_u}^2$, and $m_{H_S}^2$ are computed at the GUT scale but can vary from $m_0$, and the trilinear couplings $A_\lambda, A_\kappa$ are allowed also to differ from $A_0$. Thus, at tree-level, the behavior of the scNMSSM Higgs sector is controlled by these parameters,

$m_0, m_{1/2}, A_0, A_\lambda, A_\kappa, \lambda, \kappa, \tan\beta$ and $\mu_{eff}$

Here, the first five parameters are given at the GUT scale, while the last four parameters are electroweak scale parameters.

### Scan strategy

In this work, the NMSSM parameter space was extensively scanned with GUT boundary conditions using the NMSSMTools (v.5.5.2) package [38–42]. To date, over 70 theoretical and experimental constraints are implemented in this package. All theoretical and experimental constraints, except for $(g-2)_\mu$, were considered during this study. To study in detail the effect of current constraints on the Higgs sector in this model, we allowed the passage of points with problems related to the limits imposed on the Higgs while we ruled out the other ones. After that, we analyzed the points and divided them into categories according to the problems that were excluded because of them. All these points were represented in different colors. Then, we investigated the possibility of observing one of the lightest neutral Higgs bosons ($h_1, h_2$, and $a_1$) for the surviving points. It is worth mentioning that the constraints from ATLAS and CMS are taken into account represent data of proton-proton collisions at $\sqrt{s} = 13$ TeV, corresponding to the luminosity of $\sim$ 36 $fb^{-1}$ and $\sim$ 80 $fb^{-1}$ [43–55].

In a previous study, we showed the effect of the recent constraints on the properties of the neutral Higgs bosons, such as the mass spectrum, the singlet and doublets components, the reduced couplings of such particles, and their decays and reduced cross-section to bosons [37]. This paper will show the rest of the results of the possible decays of these particles and the reduced cross-section of them to fermions. Note that the reduced cross-section represents an insightful approximation of the signal strength. It is computed from the multiplication of the relevant coupling squared (e.g. effective coupling of the Higgs to gluons) relative to the SM with the branching ratio of the Higgs decays to XX in the NMSSM relative to the SM, and given by the following expression (see for e.g. [21])

$$R_{XX} = C^2(H_i) \times R_{XX}^{BR}(H_i) \quad (7)$$

The NMSSMTools package provides the reduced cross-section for each decay mode for all Higgs bosons.

Because there are many free parameters for the NMSSM, in order to simplify our analysis, we assumed the scNMSSM. With this assumption, we randomly scan scanned up to $10^8$ points to get a complete picture of this model. Moreover, we ensure that $\lambda$ at the SUSY scale is smaller than < 0.7 to avoid any Landau singularities below the GUT scale. The free parameters ranged between the following values:

500 $GeV \leqslant m_0, m_{1/2}$ (GUT) $\leqslant 4000$ $GeV$, 0.4 $\leqslant \lambda$ (SUSY) $\leqslant 0.7$,
$-3000$ $GeV \leqslant A_0, A_\lambda, A_\kappa$ (GUT) $\leqslant 3000$ $GeV$,
0.3 $\leqslant \kappa$ (SUSY) $\leqslant 0.7$,
100 $GeV \leqslant \mu_{eff}$ (SUSY) $\leqslant 1500$ $GeV$,
1 $\leqslant \tan\beta$ ($M_Z$) $\leqslant 10$.

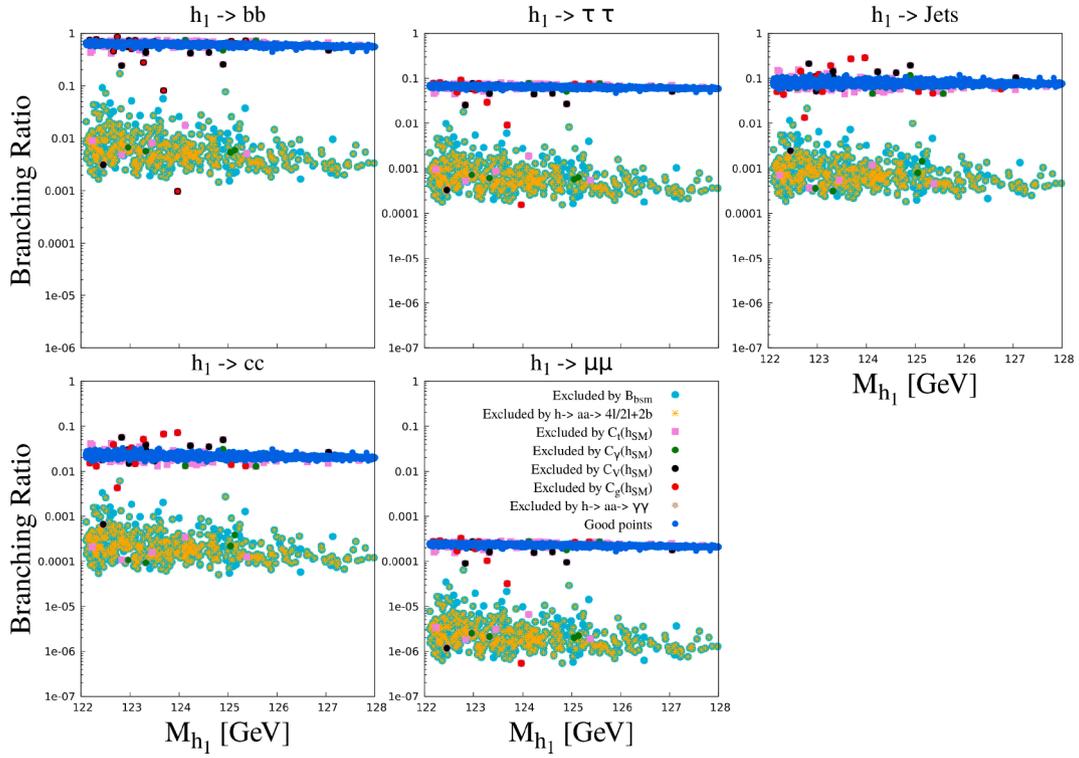

**Fig. 1.** The branching ratios of $h_1 \to bb$, $h_1 \to \tau\tau$, $h_1 \to Jets$, $h_1 \to cc$, and $h_1 \to \mu\mu$ plotted against the lightest CP-even Higgs mass $M_{h_1}$ for $A_0 \neq A_\lambda \neq A_\kappa$.

## Results

### Branching Ratios

#### Lightest CP-even Higgs ($h_1$)

Fig. 1 shows the allowed fermionic decays of $h_1$ as a function of their mass. There are some excluded points because the values of the Higgs couplings to top quarks, photons, vector bosons, gluons ($C_t$, $C_\gamma$, $C_V$, $C_g$) and the branching ratio of the SM-like Higgs to new physics ($B_{BSM}$) are 2 $\sigma$ away from the practically measured values, which represented by violet, green, black, red, and cyan colors, respectively. While the limitations from $h \to aa \to 4l/2l+2b$ and $h \to aa \to \gamma\gamma$ resulted in the exclusion of orange and red points, respectively. Finally, the blue dots express the surviving points that exceeded all recent constraints. Please note that the deviations from the SM predictions are computed with the assumption that there is only one underlying state at 125 GeV ($h_{SM}$). According to the results shown in this figure, this boson represents the SM-like Higgs with a mass ranging from 122 GeV to 128 GeV. Therefore, the branching ratios and the reduced cross-sections of this particle are supposed to be close to what was expected in the SM. As seen from this figure (in the top panels), for the allowed range mass, the branching ratio of $h_1 \to bb$, $h_1 \to \tau\tau$, and $h_1 \to Jets$ may vary from 0.4 to 0.7, 0.05 to 0.08, and 0.06 to 0.12, respectively. Next, the bottom panels shows Br ($h_1 \to cc$) runs from 0.02 to 0.03, while for $h_1 \to \mu\mu$, the allowed band indicates the expected branching ratio at 0.00025. This figure also shows that some points violate the LHC constraints. The constraints with the most considerable impact are $h \to aa \to 4l/2l+2b$ and $B_{BSM}(h_{SM})$, with the branching ratios below the accepted value in the SM. It should be noted that, in the scanned parameter space, if a region contains both good points and ruled-out points, then we plot the good points on top of the bad points since such a region is still valid given the presence of good solutions. However, if only bad points are shown in the plot, then this means we could not find good solutions in the associated region.

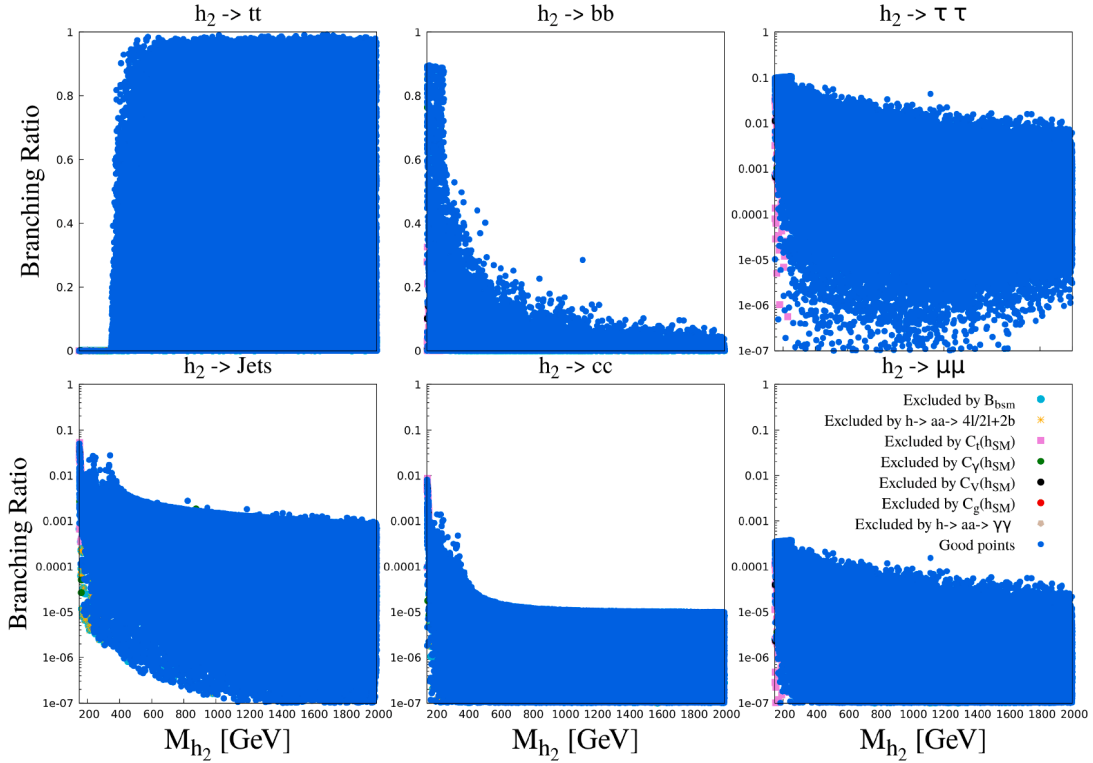

**Fig. 2.** The branching ratios of $h_2 \to tt$, $h_2 \to bb$, $h_2 \to \tau\tau$, $h_2 \to Jets$, $h_2 \to cc$, and $h_2 \to \mu\mu$ plotted against the next to lightest CP-even Higgs mass $M_{h_2}$ for $A_0 \neq A_\lambda \neq A_\kappa$.

*Second lightest CP-even Higgs ($h_2$)*

Fig. 2 shows the permitted fermionic decays of the second lightest CP-even Higgs boson, $h_2$. The results illustrate that the dominant decay mode is to $tt$ for mass values above 300 GeV for this boson. On the other hand, when the Higgs mass below 300 GeV, $h_2 \to bb$ can be dominant with a branching ratio of about 0.9. The branching ratio of the channels $h_2 \to \tau\tau, h_2 \to Jets$ are about 0.1 for $m_{h_2} < 300$ GeV, then decreases to 0.01 and 0.001, respectively, as $m_{h_2} \to 2$ TeV. For $m_{h_2}$ below 200 GeV, the branching ratio of $h_2$ is to $cc$ is around 0.01, then sharply decreases to $10^{-5}$. Finally, for the low mass range ($m_{h_2} < 200$ GeV), the value of the Higgs's branching ratio to $\to \mu\mu$ can not exceed 0.001. Moreover, the region where $m_{h_2} \leqslant 250$ GeV and branching ratios below 0.01, the impact of the constraints are more visible due to violating the constraint on $C_t(h_{SM})$ (as clearly shown for $h_2 \to \tau\tau/\mu\mu$), $B_{BSM}, C_\gamma(h_{SM})$, and from

$h \to aa \to 4l/2l+2b$ (as shown for $h_2 \to Jets/cc$).

*Lightest CP-odd Higgs ($a_1$)*

The branching ratios of allowed fermionic decays of $a_1$ are presented in Fig. 3. As shown in the right top panel, with $m_{a_1}$ above 350 GeV, the dominant decay mode is to $tt$. For $m_{a_1} < 400$ GeV, the branching ratio of $a_1 \to bb$ reaches a maximum value of 0.9 then decreases with mass increasing to about 0.45. Moving now to $a_1 \to cc$, when the mass below 400 GeV, this decays also has a maximum value of 0.01 then sharply drops to about $10^{-5}$ for the allowed range mass. Next, The decay of $a_1 \to Jets$ is possible for mass below 350 GeV with branching ratios dramatically increases from 0 to about 0.6 at $m_{a_1} \sim 350$ GeV. Finally, for the allowed range mass of the lightest CP-odd, $m_{a_1}$, the expected value of the branching ratios to $\tau\tau$ and $\mu\mu$ can reach about 0.1 and $1.5 \times 10^{-4}$, respectively. There are some points that are excluded due to the LHC constraints on $C_t(h_{SM})$ and $C_\gamma(h_{SM})$, with $m_{a_1} < 300$ GeV and BR below $10^{-5}$ (as shown for $a_1 \to cc/\mu\mu$).

*Reduced cross-sections*

Fig. 4 shows the reduced cross-section of the lightest CP-even and CP-odd Higgs bosons ($h_1, h_2$, and $a_1$) into bottom quarks via ttH production mode (top panels), and via VBF and VH production modes (bottom panels). While the reduced cross-sections of the these neutral Higgs bosons into $\tau\tau$ via ggF production mode (top panels) and via VBF and VH production modes (bottom panels) are presented in Fig. 5.

As we mentioned before, the results showed that the SM-like Higgs is the lightest of CP-even Higgs boson during this range of the scan. Therefore, the left panels in Figs. 4 and 5 show that the permissible values for the reduced cross-section of $h_1$ into $bb$ and $\tau\tau$ via various production modes are near unity. These plots also show that when the cross-section values are between 0.1 and 0.0012, there are points that were excluded due to their violation of the restrictions of LHC (specifically from $B_{BSM}(h_{SM})$ and $h \to aa \to 4l/2l + 2b$), while near the permissible values there are points that were excluded due to the constraints on $C_t(h_{SM}), C_g(h_{SM})$, and $C_V(h_{SM})$.

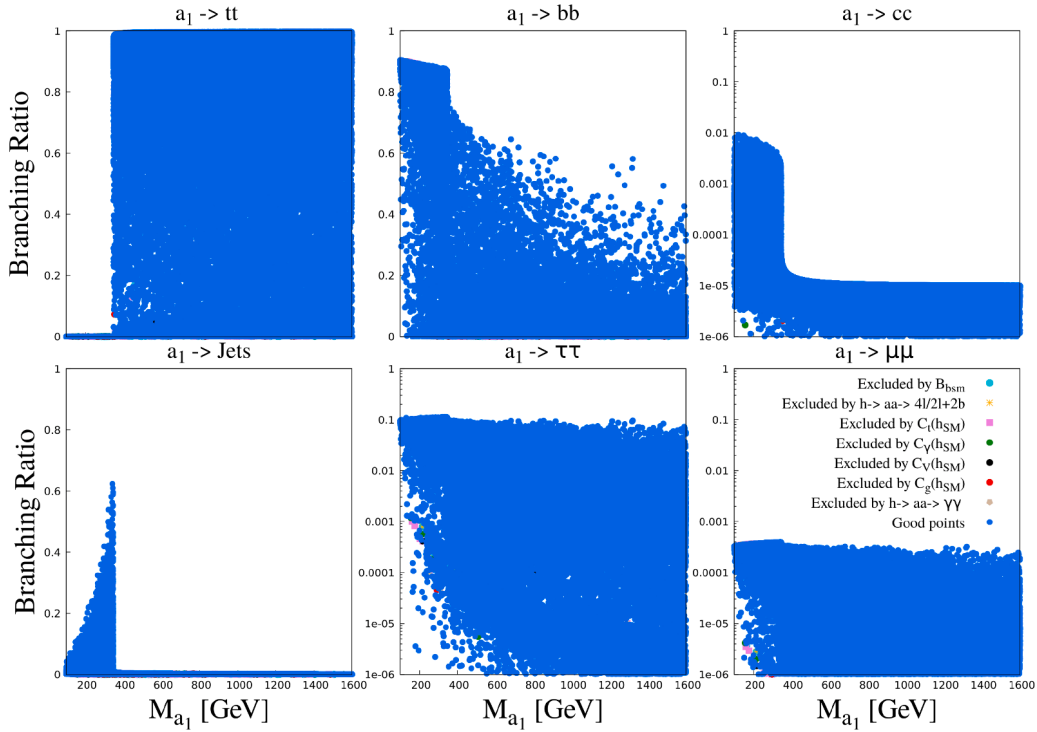

**Fig. 3.** The branching ratios of $a_1 \to tt, a_1 \to bb, a_1 \to cc, a_1 \to Jets, a_1 \to \tau\tau$, and $a_1 \to \mu\mu$ plotted aginest the lightest CP-odd Higgs mass $M_{a_1}$ for $A_0 \neq A_\lambda \neq A_\kappa$.

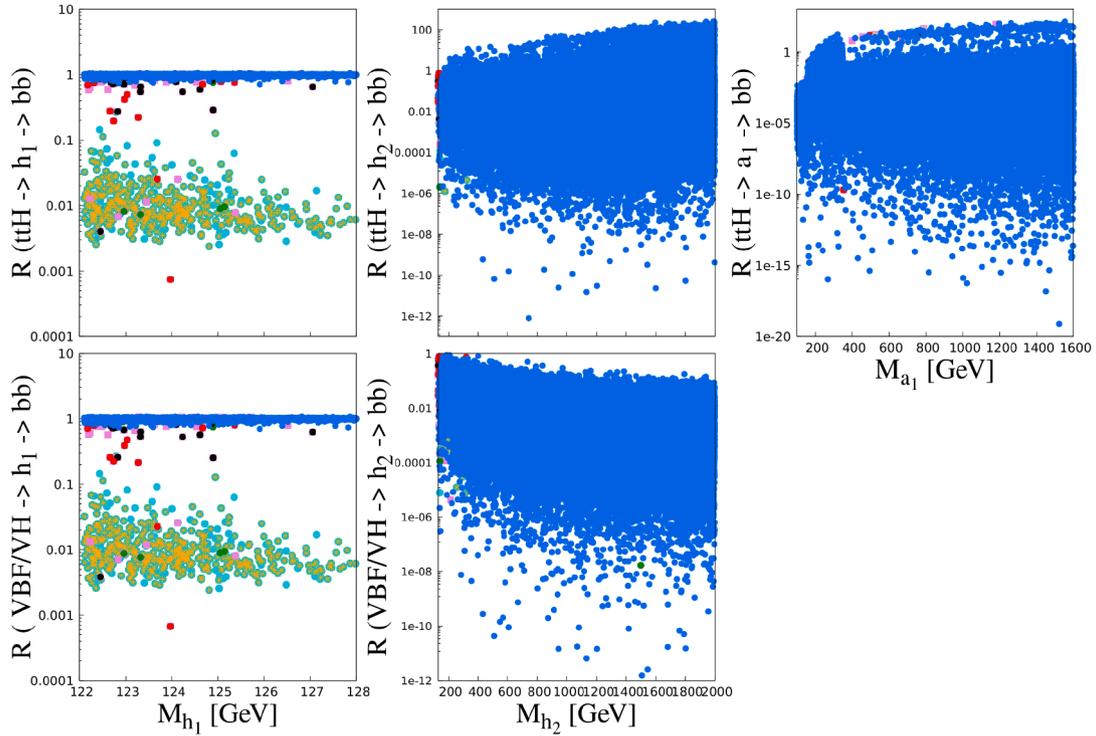

**Fig. 4.** The reduced cross-section of the lightest CP-even Higgs boson (left panels), the next to lightest (middle panels) CP-even Higgs boson, and the lightest CP-odd Higgs boson (right panel) into *bb* via ttH, VBF, and VH production modes. The color-coding is the same as Fig. 3.

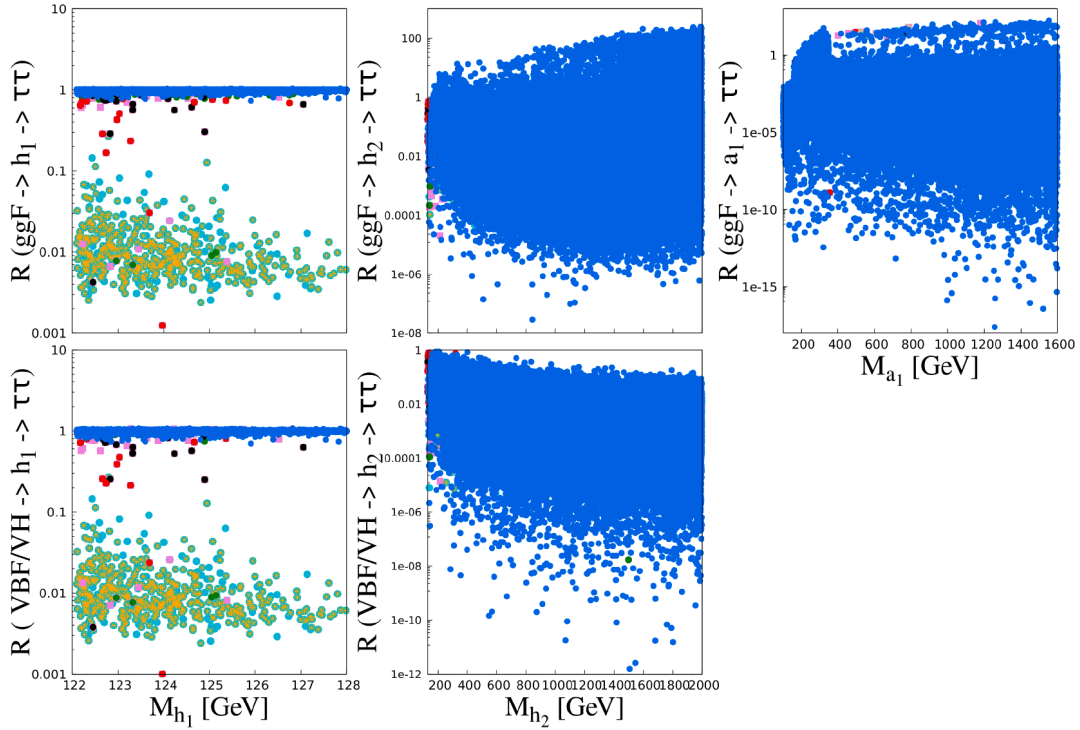

**Fig. 5.** The reduced cross-section of the lightest CP-even Higgs boson (left panels), the next to lightest (middle panels) CP-even Higgs boson, and the lightest CP-odd Higgs boson (right panel) into $\tau\tau$ via ggF, VBF, and VH production modes. The color-coding is the same as Fig. 3.

From the middle panels of Figs. 4 and 5, the results show that when the mass of $h_2 < 200$ GeV, the reduced cross-section into *bb* and $\tau\tau$ is around the unity. With mass increasing, the reduced cross-section may get enhanced by over 100 via ttH production mode and get decreased to below 0.5 via VBF or VH. We noted that when the reduced cross-section is between $10^{-4}$ and $10^{-6}$, some points were excluded due to the constraints from $h \to aa \to 4l/2l + 2b$, $B_{BSM}(h_{SM})$, $C_\gamma(h_{SM})$, and $C_t(h_{SM})$. Moreover, $R(ggF \to h_2 \to \tau\tau)$ and $R(VBF/VH \to h_2 \to \tau\tau)$ have similar properties of $R(ttH \to h_2 \to bb)$ and $R(VBF/VH \to h_2 \to bb)$, respectively.

The reduced cross-section of $a_1$ into *bb*, via ttH, and $\tau\tau$, via VBF or VH, is presented in the right panel in Figs. 4 and 5, respectively. As shown here, when the mass of $a_1 < 200$ GeV, the reduced cross-section to down-type fermions via different production modes has tiny values, and whenever the mass increases, these values get increased. Finally, We also noted that when the mass of $a_1$ above 400 GeV, there are excluded points, with huge value of the reduced cross-section, due to the limitations on $C_t(h_{SM}), C_g(h_{SM}), C_V(h_{SM})$, and from $h \to aa \to \gamma\gamma$.

## Conclusion

For the lightest neutral Higgs bosons in the NMSSM ($h_1, h_2$, and $a_1$), we examined the reduced cross-section and possible decay modes to the fermions in a region where $\lambda$ is large as possible, and $\tan\beta$ is small. The results show that in future searches the following channels could be most promising. For example, the decays of $h_2 \to b\bar{b}$, and $a_1 \to b\bar{b}$, can be dominant with a branching ratio of 0.9 when the Higgs mass below 300 GeV and 400 GeV, respectively, due to the high values of the doublet components of $h_2$ and $a_1$. Thus, the coupling between these Higgs states and $b$ quarks will get enhanced. Moreover, the branching ratio of $h_2$ and $a_1$ to a pair of top quarks, if kinematically allowed, would be dominant. This is because the doublet component of $h_2$ and singlet component of $a_1$ have high values, which leads to a strong coupling between these particles and the top quarks. The results also showed a similar behavior of a reduced cross-section of the neutral Higgs bosons into $b\bar{b}$ and $\tau\bar{\tau}$ final states.

In summary, the discovery of a SM-like Higgs boson with a mass of about 125 GeV may indicate the existence of extended Higgs sectors predicted by new physics models, such as the NMSSM. The measured couplings of the SM-like Higgs can indirectly affect the parameter space of the other non-SM Higgs bosons in the model as is shown by the findings of this paper, which can provide a deeper insight into the model and suggest new directions for future research.

## Declaration of Competing Interest

The authors declare that they have no known competing financial interests or personal relationships that could have appeared to influence the work reported in this paper.

## Acknowledgement

The authors would like to thank the Deanship of Scientific Research in King Saud University for funding and supporting this research through the initiative of DSR Graduate Students Research Support (GSR).

## References

[1] Aad Georges, et al. Observation of a new particle in the search for the Standard Model Higgs boson with the ATLAS detector at the LHC. Phys Lett B 2012;716:1–29.
[2] Chatrchyan Serguei, et al. Observation of a New Boson at a Mass of 125 GeV with the CMS Experiment at the LHC. Phys Lett B 2012;716:30–61.
[3] Haber HE, Kane GL. The search for supersymmetry: Probing physics beyond the standard model. Phys Rept 1985;117:75–263.
[4] Tanabashi M, et al. Review of Particle Physics. Phys Rev D 2018;98(3):030001.
[5] Barbieri Riccardo, Giudice GF. Upper Bounds on Supersymmetric Particle Masses. Nucl Phys B 1988;306:63–76.
[6] Kitano Ryuichiro, Nomura Yasunori. Supersymmetry, naturalness, and signatures at the LHC. Phys Rev D 2006;73:095004.
[7] Kim Jihn E, Nilles Hans Peter. The mu Problem and the Strong CP Problem. Phys Lett B 1984;138:150–4.
[8] Ellwanger Ulrich, Hugonie Cyril, Teixeira Ana M. The next-to-minimal supersymmetric standard model. Phys Rep Nov 2010;496(1–2):1–77.
[9] Maniatis M. The Next-to-Minimal Supersymmetric extension of the Standard Model reviewed. Int J Mod Phys A 2010;25:3505–602.
[10] Ellis John R, Gunion JF, Haber Howard E, Roszkowski L, Zwirner F. Higgs Bosons in a Nonminimal Supersymmetric Model. Phys Rev D 1989;39:844.
[11] Drees Manuel. Supersymmetric Models with Extended Higgs Sector. Int J Mod Phys A 1989;4:3635.
[12] Ellwanger Ulrich. A Higgs boson near 125 GeV with enhanced di-photon signal in the NMSSM. JHEP 2012;03:044.
[13] Gunion John F, Jiang Yun, Kraml Sabine. The Constrained NMSSM and Higgs near 125 GeV. Phys Lett B 2012;710:454–9.
[14] King SF, Muhlleitner M, Nevzorov R. NMSSM Higgs Benchmarks Near 125 GeV. Nucl Phys B 2012;860:207–44.
[15] Jun-Jie Cao, Zhao-Xia Heng, Jin Min Yang, Yan-Ming Zhang, and Jing-Ya Zhu. A SM-like Higgs near 125 GeV in low energy SUSY: a comparative study for MSSM and NMSSM. JHEP, 03:086, 2012.
[16] Daniel Albornoz Vasquez, Genevieve Belanger, Celine Boehm, Jonathan Da Silva, Peter Richardson, and Chris Wymant. The 125 GeV Higgs in the NMSSM in light of LHC results and astrophysics constraints. Phys. Rev. D, 86:035023, 2012.
[17] Schmidt-Hoberg Kai, Staub Florian. Enhanced $h \to \gamma\gamma$ rate in MSSM singlet extensions. JHEP 2012;10:195.
[18] Benbrik R, Gomez Bock M, Heinemeyer S, Stal O, Weiglein G, Zeune L. Confronting the MSSM and the NMSSM with the Discovery of a Signal in the two Photon Channel at the LHC. Eur Phys J C 2012;72:2171.
[19] Belanger Genevieve, Ellwanger Ulrich, Gunion John F, Jiang Yun, Kraml Sabine, Schwarz John H. Higgs Bosons at 98 and 125 GeV at LEP and the LHC. JHEP 2013;01:069.
[20] Choi Kiwoon. Sang Hui Im, Kwang Sik Jeong, and Masahiro Yamaguchi. Higgs mixing and diphoton rate enhancement in NMSSM models. JHEP 2013;02:090.
[21] King SF, Mühlleitner M, Nevzorov R, Walz K. Natural NMSSM Higgs Bosons. Nucl Phys B 2013;870:323–52.
[22] Christensen Neil D, Han Tao, Liu Zhen, Shufang Su. Low-Mass Higgs Bosons in the NMSSM and Their LHC Implications. JHEP 2013;08:019.
[23] Badziak Marcin, Olechowski Marek, Pokorski Stefan. New Regions in the NMSSM with a 125 GeV Higgs. JHEP 2013;06:043.
[24] Moretti S, Munir S, Poulose P. 125 GeV Higgs Boson signal within the complex NMSSM. Phys Rev D 2014;89(1):015022.
[25] Kiwoon Choi, Sang Hui Im, Kwang Sik Jeong, Chan Beom Park. Light Higgs bosons in the general NMSSM. Eur Phys JC, 79(11):956, 2019.
[26] Beskidt C, de Boer W, Kazakov DI. Can we discover a light singlet-like NMSSM Higgs boson at the LHC? Phys Lett B 2018;782:69–76.
[27] Baum Sebastian, Shah Nausheen R, Freese Katherine. The NMSSM is within Reach of the LHC: Mass Correlations & Decay Signatures. JHEP 2019;04:011.
[28] King SF, Mühlleitner M, Nevzorov R, Walz K. Discovery Prospects for NMSSM Higgs Bosons at the High-Energy Large Hadron Collider. Phys Rev D 2014;90(9):095014.
[29] Cao Junjie, He Yangle, Shang Liangliang, Zhang Yang, Zhu Pengxuan. Current status of a natural NMSSM in light of LHC 13 TeV data and XENON-1T results. Phys Rev D 2019;99(7):075020.
[30] Ellwanger Ulrich. Higgs pair production in the NMSSM at the LHC. JHEP 2013;08:077.
[31] Ellwanger Ulrich, Hugonie Cyril. Higgs bosons near 125 GeV in the NMSSM with constraints at the GUT scale. Adv High Energy Phys 2012;2012:625389.
[32] Potter CT. Natural NMSSM with a Light Singlet Higgs and Singlino LSP. Eur Phys JC 2016;76(1):44.
[33] Maien Y. Binjonaid, Stephen F. King. Naturalness of scale-invariant NMSSMs with and without extra matter. Phys Rev D, 90(5):055020, 2014. [Erratum: Phys. Rev. D 90, 079903 (2014)].
[34] Aldufeery Elham, Binjonaid Maien. Dark matter constraints and the neutralino sector of the scNMSSM. Universe 2021;7(2):31.
[35] Cerdeno David G, Ghosh Pradipta, Park Chan Beom. Probing the two light Higgs scenario in the NMSSM with a low-mass pseudoscalar. JHEP 2013;06:031.
[36] Cerdeño David G, Ghosh Pradipta, Park Chan Beom, Peiró Miguel. Collider signatures of a light NMSSM pseudoscalar in neutralino decays in the light of LHC results. JHEP 2014;02:048.
[37] Marwa Telba, Maien Binjonaid. Impact of LHC Higgs couplings measurements on bosonic decays of the neutral Higgs sector in the scNMSSM. 9 2020.
[38] NMSSMTOOLS - Home, https://www.lupm.univ-montp2.fr/users/nmssm/index.html.
[39] Ellwanger Ulrich, Gunion John F, Hugonie Cyril. NMHDECAY: A Fortran code for the Higgs masses, couplings and decay widths in the NMSSM. JHEP 2005;02:066.
[40] Ellwanger Ulrich, Hugonie Cyril. NMHDECAY 2.0: An Updated program for sparticle masses, Higgs masses, couplings and decay widths in the NMSSM. Comput Phys Commun 2006;175:290–303.
[41] Das Debottam, Ellwanger Ulrich, Teixeira Ana M. NMSDECAY: A Fortran Code for Supersymmetric Particle Decays in the Next-to-Minimal Supersymmetric Standard Model. Comput Phys Commun 2012;183:774–9.
[42] Muhlleitner M, Djouadi A, Mambrini Y. SDECAY: A Fortran code for the decays of the supersymmetric particles in the MSSM. Comput Phys Commun 2005;168:46–70.
[43] Albert M Sirunyan et al. Combined measurements of Higgs boson couplings in proton–proton collisions at $\sqrt{s} = 13$ TeV. Eur. Phys. J. C, 79(5):421, 2019.
[44] Search for new resonances in the diphoton final state in the mass range between 70 and 110 GeV in pp collisions at $\sqrt{s} = 8$ and 13 TeV. 2017.
[45] Aaboud Morad, et al. Search for Higgs boson decays into pairs of light (pseudo) scalar particles in the $\gamma\gamma j j$ final state in pp collisions at $\sqrt{s} = 13$ TeV with the ATLAS detector. Phys Lett B 2018;782:750–67.
[46] Sirunyan Albert M, et al. Search for an exotic decay of the Higgs boson to a pair of light pseudoscalars in the final state with two b quarks and two $\tau$ leptons in proton-proton collisions at $sqrts = 13$ TeV. Phys Lett B 2018;785:462.
[47] Aaboud M, et al. Search for the Higgs boson produced in association with a vector boson and decaying into two spin-zero particles in the $H \to aa \to 4b$ channel in pp collisions at $\sqrt{s} = 13$ TeV with the ATLAS detector. JHEP 2018;10:031.

[48] Sirunyan Albert M, et al. Search for additional neutral MSSM Higgs bosons in the $\tau\tau$ final state in proton-proton collisions at $sqrts = 13$ TeV. JHEP 2018;09:007.
[49] Sirunyan Albert M, et al. Search for an exotic decay of the Higgs boson to a pair of light pseudoscalars in the final state of two muons and two $\tau$ leptons in proton-proton collisions at $sqrts = 13$TeV. JHEP 2018;11:018.
[50] A search for pair production of new light bosons decaying into muons at sqrt(s)=13 TeV. 2018.
[51] Aaboud Morad, et al. Search for Higgs boson decays into a pair of light bosons in the $bb\mu\mu$ final state in pp collision at $\sqrt{s} = 13$ TeV with the ATLAS detector. Phys Lett B 2019;790:1–21.
[52] Aaboud Morad, et al. Search for additional heavy neutral Higgs and gauge bosons in the ditau final state produced in 36 fb$^1$ of pp collisions at $sqrts = 13$TeV with the ATLAS detector. JHEP 2018;01:055.
[53] Aad Georges, et al. Combined measurements of Higgs boson production and decay using up to 80 fb$^{-1}$ of proton-proton collision data at $\sqrt{s} = 13$ TeV collected with the ATLAS experiment. Phys Rev D 2020;101(1):012002.
[54] Search for resonances in the 65 to 110 GeV diphoton invariant mass range using 80 fb$^{-1}$ of pp collisions collected at $\sqrt{s} = 13$ TeV with the ATLAS detector. 7 2018.
[55] Combined measurements of Higgs boson production and decay using up to 80 fb$^{-1}$ of proton–proton collision data at $\sqrt{s} = 13$ TeV collected with the ATLAS experiment. 7 2018.